\documentstyle[12pt,fleqn,epsf]{article}

\advance \topmargin by -\headheight
\advance \topmargin by -\headsep
     
\evensidemargin \oddsidemargin
\renewcommand{\d}{\partial}

\begin{document}

\title{\Large\bf Inverse meson mass ordering in color-flavor-locking
phase of high density QCD}

\author{D.T.~Son$^{1,3}$ and M.A.~Stephanov$^{2,3}$\\
{\small\em $^1$ Physics Department, Columbia University, New York, NY 10027} \\
{\small\em $^2$ Department of Physics, University of Illinois,
Chicago, IL 60607-7059}\\
{\small\em $^3$ RIKEN-BNL Research Center, Brookhaven National Laboratory,
Upton, NY 11973}}
\date{October-November 1999}
\maketitle
\vskip 1in


\begin{abstract}
We derive the effective Lagrangian for the low-energy massive meson
excitations of the color-flavor-locking (CFL) phase of QCD with 3
flavors of light quarks.  We compute the decay constants, the maximum
velocities, and the masses of the mesons at large baryon chemical
potential $\mu$. The decay constants are linear in $\mu$.  The meson
maximum velocities are close to that of sound.  The meson masses in
the CFL phase are significantly smaller than in the normal QCD vacuum
and depend only on bare quark masses.  The order of the meson masses
is, to some extent, reversed compared to that in the QCD vacuum. 
In particular, the lightest particle is $\eta'$.
\end{abstract}

\newpage
\section{Introduction}

The behavior of QCD at finite baryon density could affect the physics of
neutron stars, supernovas and of heavy-ion collisions.  Quark pairing
at the Fermi surface leading to diquark condensation and color
superconductivity is a subject of many recent theoretical
studies \cite{BailinLove}-\cite{Evans}.  In particular, as first
pointed out by Alford, Rajagopal and Wilczek \cite{cfl}, in the case
of 3 light flavors the diquark condensate can ``lock'' color and
flavor symmetry rotations.  The result is the color-flavor-locking
(CFL) phase where an interesting pattern of chiral symmetry breaking
emerges: $SU(3)_L\times SU(3)_R\times SU(3)_{\rm color}\to
SU(3)_{L+R+{\rm color}}$.  The CFL phase has many similarities with
the chiral symmetry breaking QCD vacuum and nuclear matter. This
observation has
lead to the conjecture that quark and nuclear matter might be
continuously connected \cite{continuity}.

As already pointed out in Refs.\ \cite{cfl,continuity}, the CFL phase is
characterized by a hierarchy of energy scales.  At the lowest scale
lie 10 (pseudo-)Goldstone modes, arising from the breaking of axial
flavor symmetry, baryon $U(1)_B$ symmetry, 
and axial $U(1)_A$ symmetry.  These modes, except for
the $U(1)_A$, would be massless if quark masses were zero, but in reality
they do have small masses.  At a higher scale there 
are quark excitations, which
are separated by the superconducting (BCS)
 gap $\Delta$, and the gluons which acquire mass of order
$g\mu$ from the Meissner effect.  The Goldstone modes, therefore, dominate
the physics at energy scales smaller than $\Delta$ and can be described
by an effective theory.  When quarks are massless, the Lagrangian of
such a theory can be shown to be just the QCD chiral Lagrangian,
which contains the decay constants and the velocities of the Goldstone
bosons as free parameters \cite{Z,CG}.

In this paper, we show that {\em all} parameters of the chiral
Lagrangian, including the mass term, can be completely determined in
the weak-coupling regime, or the regime of very high densities.  Our
results can be summarized as follows.  We will use the ``vacuum''
notations for the Goldstone bosons, where we have an octet of
pseudoscalars $\pi$, $K$ and $\eta$ and a singlet pseudoscalar
$\eta'$.  The Goldstone boson related to the breaking of $U(1)_B$ will
be denoted as $H$.  To the leading order in strong coupling, the decay
constants of particles in the octet and of the singlets $\eta'$ and
$H$ are linear in the chemical potential, $\mu$:
\begin{eqnarray}
  f_\pi \approx 0.209\mu \qquad \mbox{and} \qquad 
  f_H = f_{\eta'} \approx 0.195\mu .
\end{eqnarray}
The dispersion relation of the mesons has the form 
$\epsilon^2=v^2p^2+m^2$, where the maximum velocities of all 
the mesons are
close to the speed of sound in
ultrarelativistic fluids:
\begin{equation}
  v_\pi = v_H = v_{\eta'} = {1 \over \sqrt{3}}.
\end{equation}

The $H$ meson remains exactly massless, while all the others acquire masses 
when nonzero quark masses $m_q$ are taken into account.  
Due to the approximate $U(1)_A$ symmetry at large $\mu$
the meson masses are not proportional to
$m_q^{1/2}$, but to $m_q$ \cite{cfl}.  In the limit $m_u, m_d\ll m_s$, 
5 mesons ($\eta$ and the kaons) have masses of the order $m_s$, while
4 others ($\eta'$ and the pions) have masses of the order $(m_{u,d}m_s)^{1/2}$.
The masses of the charged pions and the kaons are given by:
\begin{eqnarray}\label{pikmasses}
m^2_{\pi^\pm} = C (m_u+m_d) m_s + 2 C' (m_u + m_d)(2m_u + 2m_d + m_s);
\nonumber\\
m^2_{K^0} = C (m_d+m_s) m_u + 2 C' (m_d + m_s)(2m_d + 2m_s + m_u);
\nonumber\\
m^2_{K^\pm} = C (m_u+m_s) m_d + 2 C' (m_u + m_s)(2m_u + 2m_s + m_d);
\end{eqnarray}
where $C\approx1.578$, and $C'\approx0.04216$. The states of 
$\pi^0$, $\eta$ and $\eta'$ mix and their mass matrix is complicated.
In the limit $m_s\gg m_{u,d}$ the heaviest is the $\eta$ meson,
while the $\eta'$ is the lightest.

As we shall see, the meson spectrum of the CFL phase, although similar
to that of QCD vacuum, has a certain inverse mass ordering:
in particular, the neutral kaons are lighter than the charged kaons,
and the lightest particle is the $\eta'$ singlet. The kaons would
also be lighter than the pions if it was not for a seemingly small, but
important, second term in (\ref{pikmasses}), proportional to $C'$.
We shall give a simple explanation of these facts below.


\section{The effective Lagrangian}

\label{sec:eff_L}

In this section we review the basic arguments of Ref.\ \cite{CG}.  The
ground state of the CFL phase is characterized by the following
diquark condensates%
\footnote{There is also an admixture of 6-plet condensates,
which, however, is small~\cite{6plet}.}:
\begin{equation}
  X^{ia} \sim
  \epsilon^{ijk}\epsilon^{abc}\langle\psi_L^{bj}\psi_L^{ck}\rangle^*
  \qquad \mbox{and} \qquad 
  Y^{ia} \sim
  \epsilon^{ijk}\epsilon^{abc}\langle\psi_R^{bj}\psi_R^{ck}\rangle^*,
\end{equation}
where the complex conjugation was added for convenience so that $X$
and $Y$ transform under $SU(3)_c\times SU(3)_L\times SU(3)_R$ as
(3,3,1) and (3,1,3) respectively:
\begin{equation}
  \label{xytransform}
  X\to U_L X U_c^T \mbox{ and } Y\to U_R Y U_c^T.
\end{equation}
The low-energy excitations in the CFL phase are given by the slow
rotations of the phases of $X$ and $Y$.  Therefore, we can factor out
the norm of the  condensates and consider unitary matrices $X$ and
$Y$.  Together they give us $9+9=18$ degrees of freedom.  8 of
them are eaten by the gluons through the Higgs mechanism, and the
surviving 10 become low-energy excitations.

One of the Goldstones corresponds to the spontaneous breaking of the
$U(1)_A$ symmetry, which, being a symmetry of the Lagrangian, is
violated at the quantum level by the axial anomaly, or, equivalently,
by instanton-induced interactions.  
This violation, however, becomes very small at large
$\mu$ due to the effect of screening which suppresses the instanton
density by a high power of $1/\mu$ 
\cite{instantons_at_mu}.  Therefore, we can treat the
$U(1)_A$ on equal footing with
other global symmetries in this regime.

For simplicity, we shall at first ignore the overall $U(1)$ phases of
$X$ and $Y$ and assume, as in Ref.\ \cite{CG}, $\det X=\det Y=1$.  The
Lagrangian should be symmetric under the $SU(3)_c\times SU(3)_L\times
SU(3)_R$ rotations (\ref{xytransform}).  
This condition fixes the Lagrangian to the leading (second)
order in derivatives:
\begin{equation}
{\cal L}_{\rm eff} = {f_\pi^2\over2}{\rm Tr} \left[
(X^\dagger\d_0 X)^2
+  (Y^\dagger\d_0 Y)^2
\right] + \mbox{spatial gradients}
  \label{L_eff_noA}
\end{equation}
The spatial gradients enter in a similar way but with a different
constant instead of $f_\pi$.  The cross term ${\rm Tr}(X^\dagger\d_0
X)(Y^\dagger\d_0 Y)$ \cite{CG} is allowed by the symmetries, 
but, as we shall
see in Sec.\ \ref{sec:fpi}, it is suppressed in weak coupling
(i.e., at large $\mu$.)  Roughly speaking, to leading order in $g^2$,
left and right quarks decouple from each other.

Since $SU(3)_c$ is a local gauge symmetry we must replace the
derivatives in Eq.\ (\ref{L_eff_noA}) by covariant derivatives,
\begin{eqnarray}\label{XYA}
  {\cal L}_{\rm eff} &=& {f_\pi^2\over2}{\rm Tr} \left[
  (X\d_0 X^\dagger - gA_0)^2
  +  (Y\d_0 Y^\dagger - gA_0)^2
  \right] +\ldots
  \nonumber\\
  &&=
  {f_\pi^2\over4}{\rm Tr} \left[
  (X\d_0 X^\dagger - Y\d_0 Y^\dagger)^2 
  +
  (X\d_0 X^\dagger + Y\d_0 Y^\dagger - 2gA_0)^2
  \right]
  + \ldots
\end{eqnarray}

The second term is responsible for the Higgs effect: the vector-like
fluctuations, $dX=dY$, of $SU(3)$ phases become
longitudinal components of the gluon $A_\mu$.  The gluons acquire a
mass of order $O(gf_\pi)$ (electric and magnetic masses
are, in general, different).  We shall see that $f_\pi\sim\mu$, and
thus the
gluon mass is much larger than the momentum scales $p$ that we are
considering ($p<2\Delta\ll g\mu$), so gluons decouple from the
low-energy theory. The axial-like fluctuations of the phases, $d
X= -d Y$, can be written as fluctuations of the phases of a new
unitary matrix:
\begin{equation}
  \Sigma = XY^\dagger,
  \label{Sigma_def}
\end{equation}
and the effective Lagrangian takes the form (in Euclidean space):
\begin{equation}
  {\cal L}_{\rm eff} = 
  {f_\pi^2\over4}{\rm Tr} \left[
  \d_0\Sigma\d_0\Sigma^\dagger
  +
  v_\pi^2\d_i\Sigma\d_i\Sigma^\dagger
  \right].
\end{equation}
which is the usual Lagrangian of the nonlinear sigma model except that
the speed of the mesons, $v_\pi$, can
be different from the speed of light.  The
matrix $\Sigma$ is a singlet in color and transforms under $SU(3)_L\times
SU(3)_R$ as
\begin{equation}\label{sigmatransform}
  \Sigma\to U_L\Sigma U_R^\dagger,
\end{equation}
and describes the meson octet.

What happens if we take into account the $U(1)$ phases of $X$ and $Y$?
Then it is possible to add into the chiral Lagrangian a term
proportional to 
\begin{equation}\label{oziterm}
  ({\rm Tr} X^\dagger\d_0X)^2 + ({\rm Tr}Y^\dagger\d_0Y)^2.
\end{equation}
It is possible to add a cross term, $({\rm Tr} X^\dagger\d_0X)({\rm
Tr} Y^\dagger\d_0Y)$ but it is suppressed in weak coupling by the same
reason by which the cross term was omitted in Eq.\ (\ref{L_eff_noA}).
We shall make the consequences of the term (\ref{oziterm})
 clear by splitting $X$ and $Y$ into $SU(3)$ and $U(1)$ parts,
\begin{equation}
  X = \tilde{X} e^{ 2i \theta + 2i \phi}
  \quad\mbox{and}\quad
  Y = \tilde{Y} e^{- 2i \theta + 2i \phi};
\end{equation}
where $\tilde{X}$, $\tilde{Y}$ are $SU(3)$ matrices, and the angle
$\phi$ is the variable conjugate to the
baryon charge, normalized to 1 for a single quark.  The normalization
of the $U_A(1)$ phase $\theta$ is fixed analogously.  Consequently,
the field $\Sigma$ defined in Eq.\ (\ref{Sigma_def}) now has the form
\begin{equation}
  \Sigma = \tilde\Sigma\ e^{4i\theta}, \qquad
  \tilde\Sigma = \tilde{X}\tilde{Y}^\dagger \ .
\end{equation}
In terms of $\tilde{\Sigma}$, $\phi$ and $\theta$, the lowest order
chiral Lagrangian, consistent with the symmetries, has the form
\begin{eqnarray}
{\cal L}_{\rm eff} &=&
{f_\pi^2\over4}{\rm Tr} \left[
\d_0\tilde\Sigma\d_0\tilde\Sigma^\dagger
+
v_\pi^2\d_i\tilde\Sigma\d_i\tilde\Sigma^\dagger
\right]
+
12f_{\eta'}^2
\left[(\d_0\theta)^2 + v_{\eta'}^2(\d_i\theta)^2\right]
\nonumber\\&&
+
12 f_H^2  \left[(\d_0\phi)^2 + v_H^2(\d_i\phi)^2\right].
  \label{chiral_L_10}
\end{eqnarray}
The difference between the decay constants of $\eta'$ and $H$ mesons
from $f_\pi$ arises from the additional allowed term
(\ref{oziterm}), while the difference between $f_{\eta'}$ and
$f_H$ arises from the cross term $({\rm Tr} X^\dagger\d_0X)({\rm Tr}
Y^\dagger\d_0Y)$, and, therefore, is small at weak coupling.  We shall
use $f_{\eta'}=f_H$ in the rest of the paper.

The elementary meson fields $\pi^A$, $\eta'$, and $H$, are defined as
\begin{equation}
  \tilde{\Sigma} = \exp \biggl( i{\lambda^A\pi^A\over f_\pi} \biggr), 
  \qquad \theta = {\eta'\over \sqrt{24}f_{\eta'}}, \qquad
  \phi = {H\over \sqrt{24}f_H},
\end{equation}
where $\lambda^A$ ($A=1\ldots8$) are Gell-Mann matrices normalized so
that ${\rm Tr}\lambda^A\lambda^B=2\delta^{AB}$.

\section{Decay constants of the Goldstone bosons}

\label{sec:fpi}

Let us now show that all the decay constants of the Goldstone bosons,
$f_\pi$, $f_{\eta'}$, and $f_H$, can be computed in the high-density
regime and are all proportional to the chemical potential.  We will
demonstrate our method on $f_H$, whose calculation is the simplest.
Let us imagine that baryon symmetry is not a global symmetry, but
instead a local one.  So we introduce into the theory a gauge field
$A_\mu$ coupled to the baryon current.  Due to the Higgs mechanism,
the gauge bosons acquire a finite mass, which can be computed in both
the microscopic theory, where it is expressed via the chemical
potential, and in the effective theory, where it is proportional to
$f_H$.  Comparison of the two results leads to the determination of
$f_H$.

In order to compute $f_H$ we have to deal only with the part of the
effective chiral Lagrangian (\ref{chiral_L_10}) that contains the
phase $\phi$.  Since $\phi$ is not neutral with respect to the baryon
symmetry, which is now local, one should replace the
derivatives by covariant ones,
\begin{equation}
  \label{chiral_D}
  {\cal L} = 12 f^2_H \left[(\partial_0 \phi + eA_0)^2 +
                v_H^2 (\partial_i \phi + eA_i)^2\right],
\end{equation}
where $e$ is some arbitrary small coupling constant.  From Eq.\
(\ref{chiral_D}) we find that the gauge bosons acquire finite masses,
which are different for $A_0$ and $A_i$,
\begin{eqnarray}
  m^2_{A_0} & = & 24 e^2 f_H^2
    \label{m0H_eff} \\
  m^2_{A_i} & = & 24 v_H^2 e^2 f_H^2 \label{miH_eff}
\end{eqnarray}

On the other hand, the masses in Eqs.\ (\ref{m0H_eff},\ref{miH_eff})
can be computed from the microscopic theory, where $m^2_{A_0}$ has the
meaning of the Debye mass (or the inverse Thomas-Fermi screening
length) of the electric $A_\mu$ field, while $m^2_{A_i}$ is the Meissner
mass (or the inverse London penetration depth) of the magnetic
components of $A_\mu$.  To compute these masses, it is most convenient to
use a low-energy effective theory containing only fermion modes near
the Fermi surface, similar to the one derived by Hong \cite{Hong}.  We
shall work in Euclidean space where the action has the form
\begin{equation}
  S = \int\!{d^4p\over(2\pi)^4}\, \biggl(
      \psi^\dagger(p)(ip_0 + \epsilon_p)\psi(p) +
       e\psi^\dagger(p)\psi(p) (iA_0(0)+v_iA_i(0))\biggr) +
       \int\!d^4x\,{m_0^2\over2} A_iA_i
  \label{Hong}
\end{equation}
where $v_i=p_i/|{\bf p}|$, and we have kept only the gluon modes with
zero momentum in the interaction term since only they are needed in
future discussion.  By using the effective Lagrangian (\ref{Hong}) one
can avoid dealing with the Dirac structure of the quark propagator,
which is somewhat complicated in the superconducting phase.  The
``bare Meissner mass'' term proportional to $m_0^2$ in Eq.\
(\ref{Hong}) emerges from integration out of all degrees of
freedom except the ones near the Fermi surface; its magnitude can be
found from the following simple argument.  In the first-quantization
picture, the Hamiltonian of a relativistic particle is simply 
${\cal H}=|{\bf p}|$.  
As we couple the particle to the gauge field $A_\mu$, its
Hamiltonian becomes
\begin{equation}
  {\cal H} = |{\bf p} + e{\bf A}| +ieA_0 = 
      |{\bf p}| + ieA_0 + e({\bf v}\cdot{\bf A}) +
      {e^2 \over 2|{\bf p}|} \left({\bf A}^2 - ({\bf v}\cdot {\bf A})^2\right)
+\ldots
  \label{1st_quant}
\end{equation}
In the second quantization language, the first three terms in Eq.\
(\ref{1st_quant}) correspond to the first term of the Lagrangian
(\ref{Hong}).  The last term, if we sum over all particles with
momentum ${\bf p}$ inside the Fermi sphere, reproduce the bare
Meissner mass term in the effective Lagrangian (\ref{Hong}) with
\begin{equation}
  m_0^2 = 6e^2\,{\mu^2\over 2\pi^2}
  \label{bareMeissner}
\end{equation}
A consistency check for the Lagrangian (\ref{Hong}) is the
computation of the physical Meissner mass in the normal phase.  It can
be shown that the bare value (\ref{bareMeissner}) is exactly the one
required for the Meissner effect to be absent in the normal phase.

In the CFL phase, the quark propagator has its simplest form in the
basis
\begin{equation}\label{basis}
  \psi_{ai} = \sum_{A=1}^9 {\lambda^A_{ai}\over\sqrt{2}} \psi^A
\end{equation}
where $\lambda^A$ are Gell-Mann matrices if $A=1\ldots8$ and
$\lambda^9=\sqrt{2/3}$.  The Nambu-Gorkov quark propagators are
diagonal in this basis,
\begin{eqnarray}\label{propagators}
  \langle \psi^A(p)\psi^B(-p) \rangle & = &
    {\delta^{AB}\Delta^A \over p_0^2 + \epsilon_p^2 + \Delta_A^2};\\
  \langle \psi^A(p)\psi^{\dagger B}(-p) \rangle & = &
    {\delta^{AB}(ip_0+\epsilon_p)\over p_0^2 + \epsilon_p^2 + \Delta_A^2};
\end{eqnarray}
where $\epsilon_p=|{\bf p}|-\mu$ is the energy relative to the Fermi
surface.  The gaps $\Delta_A$ with $A=1\ldots8$ are equal, but
different from $\Delta_9$.  In weak coupling, if we denote
$\Delta_1=\cdots=\Delta_8=\Delta$, then
$\Delta_9=-2\Delta$.\footnote{Our $\Delta^A$ with $A\neq9$ and
$\Delta^9$ corresponds to $\Delta_8$ and $\Delta_1$ in the notation of
Ref.\ \cite{cfl}.}

\begin{figure} \centering
$$
      \def\epsfsize #1#2{0.5#1}
      \epsfbox{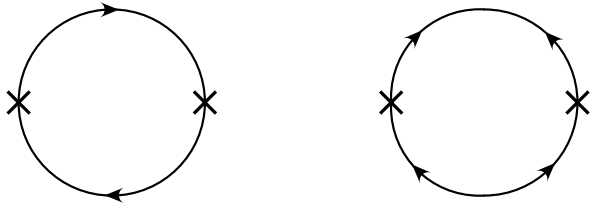}
$$
\caption{The leading order diagrams contributing to decay constants.}
\label{fig:oneloop}
\end{figure}
To the leading order, the only contribution to $m^2_{A_0}$ is from the
sum of two one-loop diagrams as shown in Fig.\ (\ref{fig:oneloop}),
which, for zero external momentum, is equal to
\begin{equation}
  2{\mu^2\over2}\int\!{dp_0\,d\epsilon_p\over(2\pi)}\, 
  \sum_{A=1}^9 \biggl(
  -{(ip_0+\epsilon_p)^2 \over (p_0^2+\epsilon_p^2+\Delta_A^2)^2} +
  {\Delta_A^2 \over (p_0^2+\epsilon_p^2+\Delta_A^2)^2} \biggr),
  \label{int_sum}
\end{equation}
where we use the formula for the phase space near the Fermi surface,
$\int{d^4p\over(2\pi)^4}={\mu^2\over2\pi^2}
\int\!{dp_0\over2\pi}\,d\epsilon_p$.  The overall factor of 2 in Eq.\
(\ref{int_sum}) comes from summation over left- and right-handed
quarks in the internal loop.  While the second diagram in Fig.\
(\ref{fig:oneloop}) is finite, the first, treated formally, has
logarithmic divergence.  The prescription to deal with the
divergence is known \cite{AGD}, and is essentially to perform
the integration over $p_0$ first. We find that the
two diagrams in Fig.\ (\ref{fig:oneloop}) give equal contribution to
the squared Debye mass which is given by
\begin{equation}
  m_{A_0}^2 = 18e^2 \,{\mu^2\over2\pi^2}.
  \label{m0H}
\end{equation}

This value is the same as the Debye mass that $A_\mu$ would have in the
absence of the superconductivity.  The origin of this coincidence can be
made clear by the following argument.  What we have computed is simply
the 00 component of the polarization operator of $A_\mu$, $\Pi_{00}$.
Since $A_0$ is coupled to the baryon charge $n=\psi^\dagger\psi$,
$\Pi_{00}$ has the following interpretation in the linear response
theory: the baryon charge density generated by an external 
uniform field $A_0$ is given by
\begin{equation}
  n = \Pi_{00}(0) A_0.
\end{equation}
But the uniform $A_0$ field is simply a shift of the chemical potential, or
the Fermi energy. Therefore, $\Pi_{00}$ is equal to $e^2\,\partial
n/\partial\mu$, i.e., the density of states near the Fermi surface, which
is exactly the right hand side of Eq.\ (\ref{m0H}).  It is easy to see
that $\partial n/\partial\mu$ is the same in the normal phase and
superconducting phase, provided that the gap in the latter is small.

Comparing Eq.\ (\ref{m0H}) with what one gets from the effective
theory, Eq.\ (\ref{m0H_eff}), one finds the decay constant $f_H$,
\begin{equation}
  f^2_H = {3\over4}{\mu^2\over 2\pi^2}.
  \label{fH}
\end{equation}
An important remark is in order here.  Eq.\ (\ref{fH}) tells us that
$f_H$ depends only on the chemical potential $\mu$,
but not on the gap $\Delta$.  This might seem contradicting the fact
that at $\Delta=0$ the $U(1)_B$ symmetry would be restored and no Goldstone
boson is expected in the theory.  The explanation is
that, as $\Delta$ decreases, the domain of applicability of the
effective Lagrangian $p<2\Delta$ shrinks and disappears at $\Delta=0$;
therefore the persistence of $f_H$ does not contradict the restoration
of $U(1)_B$ symmetry at $\Delta=0$.


To find the Meissner mass, we have to evaluate the same two diagrams
of Fig.\ (\ref{fig:oneloop}), where the vertices are attached to $A_i$
and $A_j$.  Instead of being equal, the two diagrams now cancel each
other, since the vertex factor of the first diagram is $v_iv_j$, while
in the second diagram it is $v_i(-v_j)$.  Therefore, the Meissner mass
in the superconducting phase is equal to the bare Meissner mass in the
effective Lagrangian (\ref{Hong}).  Compared to Eq.\ (\ref{miH_eff})
and taking into account Eq.\ (\ref{fH}), one finds that the velocity
of the $H$ boson is equal to $1/\sqrt{3}$, i.e. the speed of sound in
relativistic fluids,
\begin{equation}
  v_H^2 = {1 \over 3}.
  \label{vH}
\end{equation}
This result is not completely surprising, since the $U(1)_B$ phase
$\phi$ of the condensate is the variable conjugate to the baryon
density $\psi^\dagger\psi$, whose fluctuations give rise to the
sound.  Therefore, $H$ quanta can be considered as phonons.  But they
are not hydrodynamic phonons since they exist outside the hydrodynamic
regime (as in our case, at zero temperature where the mean free path
diverges.)  Moreover, as we shall explain, if one increases the
temperature the velocity of the Goldstone bosons decreases and becomes
zero at the critical temperature, while the speed of the hydrodynamic
sound is almost insensitive to the temperature in this range.
Therefore, the fact that the velocity of $H$ quanta (and other
Goldstone bosons, as we shall see) at zero temperature is equal to the
speed of hydrodynamic sound should be considered as a coincidence
bearing no fundamental reason.

\begin{figure} \centering
$$
      \def\epsfsize #1#2{0.5#1}
      \epsfbox{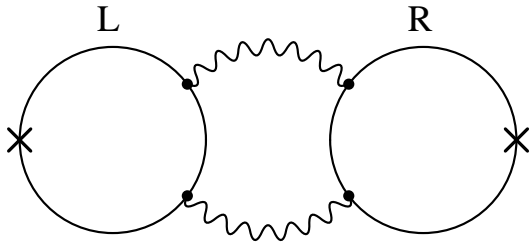}
$$
\caption{A higher-order diagram contributing to the decay constants.
One loop contains left-handed quark, the other has right-handed quark.}
\label{fig:twoloop}
\end{figure}

For the $\eta'$ meson, one can repeat the same calculation and see
that, to the leading order, all diagrams remain the same. Thus we find
that $f_{\eta'}=f_H$, and that the $\eta'$ meson also propagates with
the sound speed.  There is, however, no symmetry that requires the
decay constants of $H$ and $\eta'$ to be equal; in fact, they are not
equal in higher orders of perturbation theory.  For example, the
digram drawn in Fig.\ \ref{fig:twoloop} gives contribution of opposite
signs to $f_H$ and $f_{\eta'}$.

The computation of $f_\pi$ and $v_\pi$ is completely analogous to that
of $f_H$.  We could introduce a fictitious field coupled to the flavor
currents, but we can also make use of the existing coupling to the
gluons to compute $f_\pi$.  The only additional complication is that
the interaction vertex now has a non-trivial structure.  In our basis
(\ref{basis}), the interaction vertex has the form
\begin{equation}
  {1\over4}\sum_{A,C=1}^9\sum_{B=1}^8
  \mbox{tr}(\lambda^A\lambda^B\lambda^C)\psi^{\dagger A}A^B\psi^C
\end{equation}
Straightforward calculations show that the first diagram in Fig.\
\ref{fig:oneloop} gives $(3/4)g^2\mu^2/(2\pi^2)$ and the second
diagram contributes $-((3+4\ln2)/18)g^2\mu^2/(2\pi^2)$ to the Debye
mass.  Taking into account the factor of 2 from left- and right-handed
quarks, the result reads
\begin{equation}
  m_{A_0}^2 = {21-8\ln2\over18}{g^2\mu^2\over2\pi^2}
  \label{m0pi}
\end{equation}
The mass in Eq.\ (\ref{m0pi}) is not equal to 
the Debye mass in the normal phase, in
contrast to the previous case of the fictitious $U(1)$ boson coupled
to the baryon current.  

Notice that if the gluons are coupled to the
left-handed quarks alone, the squared Debye mass would be half
smaller, which is exactly what one obtains by throwing away $Y$ from
the Lagrangian (\ref{XYA}).  However, if we add the $XY$ cross term 
to the Lagrangian, this will no longer be true.  Therefore, this cross
term, though not forbidden by the symmetry, has a small coefficient in
weak coupling.  This can be seen also from the fact that the coupling
of the left- and right-handed flavor currents is zero at the leading
order. In higher orders of perturbation theory it receives
contribution from the diagram like in Fig.\ref{fig:twoloop}, where one
vertex corresponds to the left-handed flavor current and the other to
the right-handed one.

In the effective theory, the gluon mass is given by 
$m_{A_0}^2=g^2f_\pi^2$. Therefore,
one can determine the decay constants of the mesons in the
pseudoscalar octet,
\begin{equation}
  f_\pi^2 = {21-8\ln2\over18}{\mu^2\over2\pi^2}.
  \label{fpi}
\end{equation}
The ratio $f_\pi^2/f^2_{\eta'}=f_\pi^2/f^2_H = 2(21-8\ln2)/27\approx
1.14$ is quite close to one.  This fact seems to stem from the OZI
rule \cite{Witten}.

The Meissner mass is equal to
\begin{equation}
  m_{A_i}^2 = 2 {\mu^2\over2\pi^2} \biggl( {1\over2} - {1\over4} -
          {3+4\ln2\over54} \biggr) =
          {21-8\ln2\over54}{g^2\mu^2\over2\pi^2},
\end{equation}
where, in the parenthesis in the intermediate expression, the first
term comes from the bare Meissner mass, while the two last terms come
from the two diagrams in Fig. \ref{fig:oneloop}.  Again, in contrast
with the case of $U(1)_B$, the contribution of the two diagrams does
not cancel each other.  It is somewhat surprising that the squared
Meissner mass is $1/3$ of the squared Debye mass, as it is for the $U(1)_B$
case.  Since the ratio of these two masses is the velocity of the
Goldstone bosons, we find that the velocities of {\em all} Goldstone
modes in our theory is equal to the speed of sound $1/\sqrt{3}$
to the leading order of perturbation theory.%
\footnote{Similar results, $v=v_F/\sqrt3$, have been found by 
Bogolyubov and Anderson for phase waves in BCS superconductors and
by Leggett for spin waves in superfluid $^3$He \cite{BAL}.
}

\section{Meson masses}
\label{sec:masses}

We now turn on finite small quark masses and compute the resulting
masses of the Goldstone bosons.  By introducing bare quark masses into
the microscopic theory, we break the $SU(3)_L\times SU(3)_R$ symmetry.  
This means that more terms are allowed in ${\cal L}_{\rm eff}$.  However,
constraints on the form of these terms can still be imposed if we note
that the bare quark mass term
\begin{equation}
  \label{quarkmassterm}
  \Delta {\cal L} = \psi^\dagger_L M \psi_R + {\rm  h.c. }, 
\end{equation}
with $M$ being the $3\times3$ mass matrix, could be made invariant
under the $SU(3)_L\times SU(3)_R$ if the matrix $M$ is not passive under
this symmetry but also transforms together with $\psi_{L,R}$ in the
following way,
\begin{equation}\label{m.transform}
\psi_L\to U_L\psi_L,\quad \psi_R\to U_R\psi_R \quad
\mbox{ and } \quad M\to U_L M U_R^\dagger.
\end{equation}
Any term in the effective Lagrangian, written as a function of
$\Sigma$ and $M$, must respect the extended symmetry (\ref{sigmatransform}),
(\ref{m.transform}).  At large $\mu$ the $U(1)_A$ symmetry is
effectively restored, which imposes an additional constraint on the
possible form of mass terms in the chiral Lagrangian.  Under the
$U(1)_A$ transformation, the microscopic degrees of freedom transform
as
\begin{equation}
  \psi_L\to e^{i\zeta}\psi_L,\quad \psi_R\to e^{-i\zeta}\psi_R
  \label{psiua1}
\end{equation}
where $\zeta$ is an arbitrary pure phase.  Eq.\ (\ref{psiua1}) implies
the following transformation law of $X$, $Y$ and $\Sigma$,
\begin{equation}
  \label{sigma.eta}
  X\to e^{-2i\zeta}X,\quad Y\to e^{2i\zeta}Y \quad
  \mbox{ and } \quad \Sigma\to e^{-4i\zeta}\Sigma,
\end{equation}
The quark mass term (\ref{quarkmassterm}) is invariant under the
$U(1)_A$ symmetry if one requires that $M$ transforms as
\begin{equation}\label{m.eta}
   M\to e^{2i\zeta} M.
\end{equation}
under the $U(1)_A$.
Thus, any mass term in the effective Lagrangian must be invariant
under both the $SU(3)_L\times SU(3)_R$ and the $U(1)_A$ symmetry extended by
the transformation of $M$.%
\footnote{A similar symmetry argument allows one to 
fix the form of the effective Lagrangian in 2-color
QCD to lowest order in $\mu$ and $m_q$ \cite{KST}.}

As the bare quark masses are assumed small, we want to construct the
mass terms of lowest possible order in $M$.  To the first order, one
candidate is the mass term of the chiral Lagrangian at zero chemical
potential, ${\rm Tr} M^\dagger\Sigma$.  This term is invariant under
the $SU(3)_L\times SU(3)_R$, however it is not invariant under the
$U(1)_A$ symmetry.  Therefore, we must consider terms of higher order
in $M$.%
\footnote{The fact that $\Delta {\cal L}_{\rm eff} = O(M^2)$
means that in the chiral limit $M\to0$ the chiral condensate 
$\langle\bar\psi\psi\rangle=0$, while
$\langle(\bar\psi\psi)^2\rangle\ne0$. Such a pattern of spontaneous 
chiral symmetry breaking has been discussed in \cite{Stern}.
}

In the order $M^2$ we find that the following term is allowed by the
symmetry\footnote{ One way of looking at this term is to realize
that under $SU(3)_L$ both $M$ and $\Sigma$ transform as fundamental
3-plets. We require two powers of $M$ and one of $\Sigma$ in order to
satisfy the $U(1)_A$ neutrality.  We can construct an $SU(3)_L$
singlet out of a product of $M$, $M$ and $\Sigma$ if we antisymmetrize
with respect to the first index of each of these matrices (the index
on which $SU(3)_L$ acts).  Antisymmetrizing also with respect to the
second index we obtain an $SU(3)_L\times SU(3)_R$ singlet:
$\epsilon_{a'b'c'}\epsilon_{abc} M_{aa'}M_{bb'}\Sigma_{cc'}$, which
coincides with (\ref{detm.m-1.sigma}) up to a constant.}
\begin{equation}
  \label{detm.m-1.sigma}
  \Delta {\cal L}_{\rm eff} = 
  -c\cdot\det M\cdot {\rm Tr}(M^{-1}\Sigma) + {\rm h.c. }
\end{equation}
Another term allowed by the symmetry is%
\footnote{
The alternative linear combination of the two terms in (\ref{otherterm})
is equivalent to the term (\ref{detm.m-1.sigma}): 
$\left[{\rm Tr}(M\Sigma^\dagger)^2 - ({\rm Tr}M\Sigma^\dagger)^2\right]
 \det\Sigma = 2\det M\cdot {\rm Tr}(M^{-1}\Sigma)$.
}
\begin{equation}
\label{otherterm}
 \Delta {\cal L}_{\rm eff}' = 
- c'\left[
{\rm Tr}(M\Sigma^\dagger)^2 + ({\rm Tr}M\Sigma^\dagger)^2
\right] \det\Sigma + {\rm h.c.} 
\end{equation}

\begin{figure} \centering
$$
      \def\epsfsize #1#2{0.5#1}
      \epsfbox{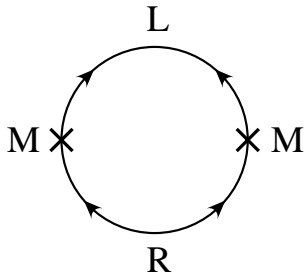}
$$
\caption{The leading order diagram contributing to the shift of the
vacuum energy by a small quark mass $M$.}
\label{fig:m2delta2}
\end{figure}

The coefficients $c$ and $c'$ of the mass terms from
(\ref{detm.m-1.sigma}) and (\ref{otherterm}) can be
calculated by matching the shift of the vacuum energy as a function
of $\Sigma$ they induce in
the effective theory: $\Delta {\cal E}_{\rm vac}(\Sigma)= \Delta {\cal L}_{\rm
eff} + \Delta {\cal L}_{\rm eff}'$, to the perturbative calculation in the
microscopic theory.  The nontrivial, $\Sigma$-dependent, shift
 in the vacuum energy due to
(\ref{quarkmassterm}) is given to the lowest order by the diagram in
Fig.\ref{fig:m2delta2}, which is similar to the second diagram
in Fig.~\ref{fig:oneloop}.
Another diagram, similar to the first diagram in Fig~\ref{fig:oneloop},
also contributes to the shift of the vacuum energy, but
its contribution does not depend on $\Sigma$.%
\footnote{This is related to the fact that
such a diagram does not depend on the values of the gaps $\Delta$
on its two fermion lines, even if they are different.} As a result,
terms such as ${\rm Tr} (M\Sigma^\dagger M^\dagger\Sigma)$
and $({\rm Tr} M\Sigma^\dagger) ({\rm Tr} M^\dagger\Sigma)$,
in principle, allowed by the symmetry,
are absent to the order we are working.

The calculation of the diagram in Fig.\ref{fig:m2delta2} gives:
\begin{eqnarray}\label{cc'}
c = {51 + 32\ln2\over108}{\mu^2\over2\pi^2} 
\qquad\mbox{ and }\qquad
c' = {15 - 16\ln2\over216}{\mu^2\over2\pi^2}.
\end{eqnarray}
The coefficient $c'$ is numerically 
very small: $c'/c\approx 0.027$, and is related to
the fact that one of the 9 gaps in the fermion propagator matrix
(\ref{propagators}) is different from the other 8.
This completely determines the mass term in the effective Lagrangian
given by (\ref{detm.m-1.sigma}) and (\ref{otherterm}).

To write the (mass)$^2$ matrix for the Goldstone bosons we expand the
field $\Sigma$:
\begin{equation}
\Sigma = \exp\left({i\pi^a\lambda^a\over f_\pi}\right) = 
1 + {i\pi^a\lambda^a\over f_\pi} - {\pi^a\pi^b\lambda^a\lambda^b\over
2f_\pi^2} + \ldots\ ,
\end{equation}
where $a=1,\ldots8,9$, $\lambda^9 = \sqrt{2/3}$ and 
$\pi^9=\eta'f_\pi/f_{\eta'}$.
Since the kinetic term is conventionally normalized we
can read off the (mass)$^2$ matrix, ${\cal M}^2_{ab}+{\cal M'}^2_{ab}$,
from the mass terms
(\ref{detm.m-1.sigma}), (\ref{otherterm}) 
(except for the trivial rescaling of the $\eta'$ field):
\begin{eqnarray}\label{mass2matrix}
{1\over2}{\cal M}^2_{ab}\pi^a\pi^b &=& {C\over2} 
\det M\cdot{\rm Tr}(M^{-1}\lambda^a\lambda^b)\pi^a\pi^b;
\nonumber\\
{1\over2}{\cal M'}^2_{ab}\pi^a\pi^b &=& C' \left\{\pi^a\pi^b\left[
({\rm Tr} M\lambda_a)({\rm Tr}M\lambda_b)
+ ({\rm Tr} M)({\rm Tr}M\lambda_a\lambda_b)\right.\right.
\nonumber\\
&&\left.\left.
+ ({\rm Tr} M\lambda_a M\lambda_b) + ({\rm Tr} M^2\lambda_a \lambda_b)
\right] + 3(\pi^9)^2 \left[({\rm Tr}M)^2 + ({\rm Tr}M^2)\right]
\right.
\nonumber\\
&&\left.
-2\sqrt6\,\pi^a\pi^9\left[({\rm Tr}M)({\rm Tr}M\lambda_a) 
+ ({\rm Tr}M^2\lambda_a)\right]
\right\};
\end{eqnarray}
where, using (\ref{fpi}) and ({\ref{cc'}),
\begin{equation}
C = {2c\over f_\pi^2} = \frac13{51+32\ln2\over21-8\ln2} \approx 1.578
\quad\mbox{ and }\quad
C' = {2c'\over f_\pi^2} = \frac16{15-16\ln2\over21-8\ln2} \approx 0.04216.
\end{equation}
Since these are pure numbers, the meson masses depend only
on the bare quark masses, in contrast to the QCD vacuum where they also
depend on the value of the chiral condensate.  

In order to understand the pattern of masses let us begin with
the first term in (\ref{mass2matrix}), since
the coefficient of this term, $C$, is much greater than that of the second
term, $C'$. 

With the quark mass matrix $M = {\rm diag}
(m_u,m_d,m_s)$, the $9\times9$ meson mass matrix (\ref{mass2matrix}) 
decomposes into a diagonal $6\times6$ matrix and a
non-diagonal $3\times3$ matrix.  The former gives rise to the mass
formulas for $\pi^\pm$, $K^\pm$, $K^0$ and $\bar K^0$:
\begin{eqnarray}
  m^2_{\pi^\pm} = C (m_u+m_d) m_s + \ldots; \quad
  m^2_{K^\pm} = C (m_u+m_s) m_d + \ldots; \nonumber\\
  m^2_{K^0} = C (m_d+m_s) m_u + \ldots\, ,
\end{eqnarray}
where ellipses denote contributions proportional to $C'$.
The remaining $3\times3$ matrix corresponding to
the $\pi^0$, $\eta$ and $\eta'$ mesons is not diagonal.  
The mixing pattern is easy to
understand if we neglect the small difference between $f_\pi$ and
$f_{\eta'}$.  Then the $3\times3$ mass matrix has the
simplest form in the basis $\bar uu$, $\bar dd$, $\bar ss$, related to
$\pi^0$, $\eta$, $\eta'$ by the quark-model relations $\pi^0=(\bar
uu-\bar dd)/\sqrt2$, $\eta=(\bar uu+\bar dd-2\bar ss)/\sqrt6$, and
$\eta'=(\bar uu+\bar dd+\bar ss)/\sqrt3$.  In this basis
$\lambda_{\bar uu}={\rm diag}(1,0,0)$, $\lambda_{\bar dd}={\rm
diag}(0,1,0)$, $\lambda_{\bar ss}={\rm diag}(0,0,1)$, and
the $3\times3$ mass matrix becomes diagonal.  The mixing between $\pi^0$,
$\eta$, and $\eta'$, thus, is ideal, and the masses of the mixed
states are
\begin{equation}
  \label{m2C}
  m_{\bar uu}^2 \cong 2C m_d m_s + \ldots, \qquad
  m_{\bar dd}^2 \cong 2C m_u m_s + \ldots, \qquad
  m_{\bar ss}^2 \cong 2C m_u m_d + \ldots.
\end{equation}
As we shall see, however, the second mass term makes a significant
correction to this ideal mixing, despite the fact that $C'\ll C$,
when $m_s\gg m_{u,d}$.

The reader will notice that the mass ordering induced by the first
term in (\ref{mass2matrix}) is
completely reversed compared to what one sees in the QCD vacuum.
While this is quite surprising, and seems unnatural under the
hypothesis of continuity between quark and hadronic matter, it can be
easily explained. The key point is that the mesons, which are
the fluctuations of $\Sigma$ in Eq.\ (\ref{Sigma_def}), should be
thought of as  bound states of a triplet antidiquark $X$ and an
antitriplet diquark $Y^\dagger$, i.e.\ as $\bar q\bar q qq$ states, rather than
$\bar qq$ states.  Consider, for example, the
$\bar ss$ state.  Now the $s$ quark should be
replaced by the $\bar u\bar d$ antidiquark, and the $\bar s$ antiquark
should be replaced by the $ud$ diquark, so this meson is represented as $\bar
u\bar d ud$.  Since such a meson does not contain the strange quark, it is
not surprising that its mass does not depend on $m_s$.  
For all other mesons, one can write the quark structure
as well, by making the replacement $u\to\bar d\bar s$, $d\to \bar
u\bar s$, $s\to \bar u\bar d$.  Since one replaces the heaviest quark
by the lightest antidiquark and vice versa, the inverse mass ordering
can be expected.\footnote{It has been suggested that a similar pattern
arises for scalar mesons in QCD vacuum \cite{jaffe}.} 
It is important to note, however, that the mesons in
the CFL phase have the same quantum numbers (up to mixing) as the
mesons in the QCD vacuum.

Although the coefficient of the second term in (\ref{mass2matrix}),
$C'$, is almost 40 times smaller than $C$, it contributes
significantly to the masses of some of the mesons, because
the ratio of $m_s$ to $m_d$ or $m_u$ is also very large.
The masses of the $\pi$ and $K$ mesons are given by (\ref{pikmasses}).
In the limit $m_s\gg m_{u,d}$ they reduce to:
\begin{equation}
m^2_{\pi^\pm} \approx C(m_u+m_d)m_s;\quad
m^2_{K^0} \approx (Cm_u + 4C'm_s)m_s; \quad
m^2_{K^\pm} \approx (Cm_d + 4C'm_s)m_s;
\end{equation}
where we treated both $C/C'$ and $m_s/m_d$
as large numbers of the same order of magnitude.
We see that in the kaon masses 
the smallness of $C'/C$ is compensated by the greatness
of $m_s/m_{u,d}$. The ordering of charged and neutral kaons remains
inverted, however.

The remaining $3\times3$ matrix for $\pi_0$, $\eta$ and $\eta'$
is rather complicated. To understand its properties it is helpful
to consider the limiting case when $m_u=m_d=0$. In this case
the matrix simplifies, and one finds that two of the three
states are massless (together with charged pions, in accordance with
the Goldstone theorem and the breaking of $SU(2)\times U(1)_A$), 
while the third has a mass of order $m_s$.
The massless states are pure $\pi^0$ and $(2\sqrt2\pi^9 - \pi^8)/3$,
which is mostly $\eta'$. 

In order to get the feeling of the numbers involved, we 
substitute for the bare quark
masses the values  $m_u=4$ MeV, $m_d=7$ MeV and $m_s=150$ MeV.
Performing numerical diagonalization of the mass matrix
(\ref{mass2matrix}) we find the
following spectrum:
$m_{\eta} = 117$ MeV, $m_{\pi^0} = 53$ MeV, $m_{\eta'} = 30$ MeV.
The mixings of $\pi^0$ with $\eta$ and $\eta'$ are on the level of a few
percent, while the mixing between $\eta$ and $\eta'$ is on the level
of 20\%. The numerical values for the masses (\ref{pikmasses})
of pions and kaons are given by: $m_{\pi^\pm} = 53$ MeV, 
$m_{K^0} = 76$ MeV, $m_{K^\pm} = 72$ MeV.

\section{Conclusion}
\label{sec:concl}

In this paper we have shown that all parameters characterizing the
dynamics of Goldstone bosons in the CFL phase, including the decay
constants, masses, and the maximum
velocities, can be reliably calculated in the
weak-coupling regime of high densities.  These parameters depend only
on very few inputs --- in fact, only the decay constants depend on the
chemical potential, while the meson masses depend only on the bare
quark masses and the velocities are given by a constant.  This fact allows
us to derive the full chiral Lagrangian without being dependent of the
calculation of, e.g., the BCS gap, whose asymptotic behavior is known
\cite{Son}, but the exact numerical value of the prefactor
is not \cite{SchaefWil1gluon,PRgap}.
 
In contrast to the meson masses in the QCD vacuum, we found that, 
when the
strange quark is much heavier than the $u$ and $d$ quarks, 5 mesons
have masses of the order of $m_s$, and 4 mesons
have masses of the order $(m_{u,d}m_s)^{1/2}$.  The lightest of those
mesons is $\eta'$ with a mass around 30 MeV.   The most
surprising fact we found is the partially inverse mass ordering of the meson
nonet.  As we explained, this can be understood if we treat the
mesons as $\bar q\bar qqq$ states, rather than $\bar qq$ states.
 
We must also discuss the region of validity of our treatment.  The
effective theory we are working with contains only meson modes,
therefore, it is applicable only for energies smaller than $2\Delta$.
In particular, a meson exists only as long as its mass is smaller than
$2\Delta$, otherwise it would rapidly decay into a particle-hole pair.
At asymptotically large chemical potential where $\Delta$ increases
with $\mu$, all mesons are stable with respect to this decay channel.
At intermediate densities, the gap may actually drop substantially
below 100 MeV in some range of the chemical potential
\cite{SchaefWil1gluon}.  If the gap is small, some of the mesons may
become unstable.  For example, for our values of the quark masses, the
$\eta$ meson disappears if $\Delta$ is smaller than about 60 MeV, and
the lightest meson, $\eta'$, disappears only when the gap drops below
15 MeV.

Another effect which becomes
important when $\mu$ is not sufficiently large 
is the explicit breaking of the $U(1)_A$
symmetry by anomalous instanton-induced interactions. This effect
gives a direct contribution (independent of the quark masses)
to the mass of the $\eta'$ state, and $m_q^{1/2}$ contributions to
the masses of the other mesons \cite{cfl}.

We can estimate the value of $\mu$ at which the transition from the
normal QCD vacuum to the CFL phase occurs very roughly by 
comparing the value of $f_\pi\approx 0.2\mu$ in the CFL phase
to its vacuum value $f_\pi=93$ MeV. We find $\mu_{\rm tr}
\sim 500$ MeV, which is reasonable and agrees with other estimates.

It is straightforward to extend our calculations to the case of finite
temperature.  Let us discuss, qualitatively, the case of
temperatures close to critical.  The Meissner masses should decrease
as the temperature increases and vanish at the critical temperature,
since one expects the Meissner effect to be absent in the normal phase.
On the other hand, Debye masses are nonzero in the normal phase, and so
should not vanish at the critical temperature.  Thus,
near the phase transition, the velocities of the Goldstone modes
become small, which is the manifestation of the critical slowing down
of the relaxation of the condensate near the phase transition.  The
decay constants remain finite, while the masses go to zero as
$\mu\Delta/T_c$ near the phase transition.  It would be 
interesting to study possible consequences of this behavior.
The inverse mass ordering and anomalous lightness of mesons,
in particular, of the $\eta'$ \cite{KKL},
in the CFL phase may also significantly affect their  production
should such phase be accessible in heavy ion collisions.

\bigskip
{\Large\bf Acknowledgements}
\bigskip

We would like to thank R. Casalbuoni, R. Gatto, 
R. Jaffe, R. Pisarski, K. Rajagopal, D. Rischke, T. Sch\"afer, I. Shovkovy
and E. Shuryak for helpful discussions and comments.


\newpage

\begin{thebibliography}{99}

\bibitem{BailinLove} D.~Bailin and A.~Love, Phys. Rept. {\bf 107}, 325
(1984), and references there in.

\bibitem{ARW} M.~Alford, K.~Rajagopal, and F.~Wilczek, Phys. Lett. B {\bf
422}, 247 (1998).
 
\bibitem{RSSV} R.~Rapp, T.~Sch\"afer, E.V.~Shuryak, and M.~Velkovsky,
Phys. Rev. Lett. {\bf 81}, 53 (1998).

\bibitem{cfl} M.~Alford, K.~Rajagopal, and F.~Wilczek, Nucl. Phys.
{\bf B537}, 443 (1999).

\bibitem{continuity} T.~Sch\"afer and F.~Wilczek,
Phys. Rev. Lett. {\bf 82}, 3956 (1999).

\bibitem{Son} D.T.~Son, Phys. Rev. D {\bf 59}, 094019 (1999).

\bibitem{SchaefWil1gluon} T.~Sch\"afer and F.~Wilczek, hep-ph/9906512.

\bibitem{PRgap} R.D. Pisarski and D.H. Rischke, nucl-th/9907041.

\bibitem{6plet} 
		 M. Alford, J. Berges, and K. Rajagopal, hep-ph/9903502;
		R.D. Pisarski and D.H. Rischke, nucl-th/9907094.

\bibitem{instantons_at_mu} 
	R. Rapp, T. Schafer, E.V. Shuryak, M. Velkovsky,
	hep-ph/9904353;
	T.~Sch\"afer, hep-ph/9909574.	

\bibitem{KST} J.B. Kogut, M.A. Stephanov, and D. Toublan,
		hep-ph/9906346; 
		J.B. Kogut, M.A. Stephanov, D. Toublan,
		J.J.M. Verbaarschot, and E. Zhitnitsky, in preparation.

\bibitem{Z}
		D.K. Hong, M. Rho, I. Zahed, hep-ph/9906551;

\bibitem{CG} 
		R.~Casalbuoni, R.~Gatto, hep-ph/9908227.

\bibitem{Hong} D.K.~Hong, hep-ph/9812510.

\bibitem{Evans} N.~Evans, S.D.H.~Hsu, and M.~Schwetz, Nucl. Phys. {\bf
B551}, 275 (1999); Phys. Lett. B {\bf 449}, 281 (1999).

\bibitem{AGD} A.A.~Abrikosov, L.P.~Gorkov, and I.E.~Dzyaloshinskii,
{\em Methods of Quantum Field Theory in Statistical Physics} 
(Prentis-Hall, 1963), Section 37.

\bibitem{Witten} E.~Witten, Ann. Phys. {\bf 128}, 363 (1980).

\bibitem{BAL}
		N.N. Bogolyubov, V.V. Tolmachev, D.V. Shirkov,
		{\it New Methods in the Theory of Superconductivity},
		New York, Consultants Bureau, 1959;
		P.W. Anderson, Phys. Rev. {\bf 112}, 1900 (1958);
		A.J. Leggett, Rev. Mod. Phys., {\bf 47}, 331 (1975). 

\bibitem{Stern}
		J. Stern, hep-ph/9801282;
		I.I. Kogan, A. Kovner, M. Shifman,
		 Phys. Rev. D {\bf 59}, 016001 (1999).

\bibitem{jaffe} R.L. Jaffe, Phys. Rev. D {\bf 15}, 267 (1977); {\bf
15}, 281 (1977).

\bibitem{KKL}
		J. Kapusta, D. Kharzeev, L. McLerran,
		Phys. Rev. D {\bf 53}, 5028 (1996).

\end{thebibliography}
\end{document}